\documentclass{cargese}\usepackage{graphicx}
\let\footnote\savefootnote
\let\footnotetext\savefootnotetext 
\let\footnoterule\savefootnoterule 
\normallatexbib

\begin{document}
\articletitle[Comments on non-supersymmetric type I vacua]{Comments on 
Non-supersymmetric type I Vacua\footnote{CPHT-PC736.0999}}

\author{K. F\"orger}

\affil{Centre de Physique Th\'eorique, \'Ecole Polytechnique\\
91128 Palaiseau Cedex, France }         
\begin{abstract}
We review open descendants of non-supersymmetric type IIB asymmetric
orbifolds with zero cosmological constant. We find that 
supersymmetry remains unbroken on the branes at all mass levels,
whereas it is broken in the bulk. 
\end{abstract}
Orbifold compactifications which break supersymmetry while
keeping the cosmological constant at one loop, two loop and 
possibly up to all orders in perturbation theory to zero,
have recently attracted much attention \cite{kks}.
A modification of the initial four dimensional
model led to a model in five dimensions,
which allows for a heterotic dual, 
thereby giving rise to non-vanishing 
non-perturbative corrections to the cosmological 
constant \cite{harv}. In the following we will focus
on the main results of the construction of open descendants for
Harvey's model, 
which has been worked out in detail in \cite{aaf} 
and using a different formalism in \cite{bg}. For a more complete
list of references and more details we refer the reader 
to \cite{aaf}.

{\it Parent closed string theory:} 
To begin with, we consider 
type II theory in $d=5$ compactified on the lattice
$\Gamma[{\rm SU}(2)]^4\oplus \Gamma_{1,1}(R)$. The radius
of $\Gamma[{\rm SU}(2)]$ is the self-dual radius
$R=\sqrt{\alpha'}$, whereas the radius of the circle of the fifth 
coordinate is left arbitrary.
Then we start modding out this theory by
the asymmetric orbifold, which is generated by
the following two elements \cite{harv}:
\begin{eqnarray*}
f &= &\left[ (-1^4 , 1 \,;\, 1^5 ) \,,\, (0^4 , v_{\rm L} \,;\, \delta^4
, v_{\rm R} ) \,,\, (-)^{F_{{\rm R}}} \right] \,,
\\
g &=& \left[ (1^5 \,;\, -1^4 ,1)\,,\, (\delta^4 ,v_{\rm R} \,;\,
0^4 , v_{\rm L} ) \,,\,(-)^{F_{{\rm L}}} \right] \,,
\end{eqnarray*}
where $\delta$ is a shift by $R/2$ and $v_{\rm L/R}$ shifts the fifth
coordinate by 
\[X_{\rm L/R}\to X_{\rm L/R}\pm \frac{1}{2}\Big(
\frac{\alpha'}{R}\pm R\Big)\ ,\]
which corresponds to a symmetric shift in 
momentum and winding modes of $\Gamma_{1,1}$,
thereby balancing the level matching condition.
The action of $(-1)^{F_{\rm L/R}}$ breaks the space-time supersymmetry
of the L/R movers and therefore the combined action of the orbifold
generators $f$ and $g$ will break supersymmetry completely.
Note that the above orbifold is non-Abelian from the space group point
of view, whereas the point group, which is obtained by modding out
the space group by pure translations, generated by $f^2$ and $g^2$, 
is an Abelian asymmetric $Z_2\times Z_2$ orbifold. This transition from the
space group to the point group modifies the lattice 
$\Gamma[{\rm SU}(2)]^4$ into the lattice $\Gamma[{\rm SO}(8)]$,
which includes a non-trivial NS-NS antisymmetric tensor $B_{ab}$ 
of rank $2$ in the internal lattice.
This fact already hints at a reduction of the resulting gauge group
of the open string spectrum \cite{toroidal,KST,carlo}.
The above orbifold can be seen as a freely acting orbifold of 
$T^4/Z_2$ by $f$. 

Acting with $Z_2\times Z_2$
on the torus partition function, we get the following massless
contributions
\begin{eqnarray*}
{\cal T}_{{\rm untw}}^{(0)} 
&\sim& |V_4 O_4 |^2 + |S_4 S_4 |^2 - (O_4 V_4 )(
\overline{C}_4 \overline{C}_4 ) - (C_4 C_4 )( \overline{O}_4
\overline{V}_4 ) \,,
\\
{\cal T}_{fg{\rm -tw}}^{(0)} &\sim & 8\; |O_4 S_4 - C_4 O_4 |^2 \,,
\end{eqnarray*}
where we used the level one  ${\rm SO}(4)$ characters 
$O_4$, $V_4$, $S_4$ and $C_4$, corresponding to the identity, vector,
spinor and conjugate spinor representation.  
From the above amplitudes we can immediately read off the spectrum
which, written in terms of five-dimensional fields, consists in: 
the metric, $7$ Abelian vectors, $6$ scalars and $8$ fermions
from the untwisted sector, and $8$ Abelian vectors, $40$ scalars
and $16$ fermions from the $fg$-twisted sector. 
The factor 8 in ${\cal T}_{fg{\rm -tw}}^{(0)}$ counts the number 
of fixed points left invariant by the shifts. 
This partition function has the remarkable property that it
is non-supersymmetric, but has the same number of fermionic
and bosonic degrees of freedom. Moreover, the torus amplitude is
invariant under T-duality and in the limit $R\to \infty$ leads to
type IIB on $T^4/Z_2$ with $21$ tensor multiplets coupled to 
${\cal N}=(2,0)$ supergravity.
  
{\it Open descendants:}
The construction of open descendants starts by adding to
$\frac{1}{2}{\cal T}$ the direct Klein bottle amplitude, which
gives at the massless level
\[
{\cal K}^{(0)} \sim \frac{1}{2}\Big[(V_4 O_4 - S_4 S_4) + \epsilon\,
(n_+-n_-)\, (O_4 S_4 - C_4 O_4)\Big] \,,
\]
where $n_+=6$ and $n_-=2$, revealing the fact that the orientifold planes
carry different charges under the $\Omega$ projection.
The two different choices $\epsilon=\pm 1$, which have recently been
discussed in \cite{ADS,carlo,AU}, give rise to a supersymmetric open string
spectrum for $\epsilon=1$ and a non-supersymmetric open string 
spectrum for $\epsilon=-1$.   
The spectrum of massless closed unoriented excitations 
then results in the following
five-dimensional fields: the metric, $2$ vectors, $5$ scalars and
$4$ fermions from the untwisted sector and 
the $fg$-twisted sector for $\epsilon=1$ contains 
 $2$ vectors, $26$ scalars and 
$8$ fermions whereas for $\epsilon=-1$ it comprises $14$ scalars, 
$6$ vectors and $8$ fermions.
For both choices the spectrum has bose-fermi degeneracy.
Since the Klein bottle amplitude only feels the left-right symmetric
part of the torus amplitude, it is supersymmetric and thus results
in a vanishing 1-loop contribution to the cosmological constant.

The open string sector is obtained by adding the direct annulus
amplitude to the direct channel M\"obius strip. 
In the following we will concentrate on the choice $\epsilon=1$.
Since the five dimensional theory is invariant under $T$-duality,
the annulus amplitude has to be parameterized by charges,
that represent a linear combination of $D9$ and $D5$ branes.
For the massless contribution to the direct annulus amplitude, we thus get:
\[
{\cal A}^{(0)}\sim \frac{1}{4}\Big\{
[I_M^2+ R_M^2] (V_4 O_4-S_4 S_4)+[I_M^2-R_M^2] 
(O_4 V_4-C_4 C_4)\Big\}\,,
\]
where $I_M$ denotes the sum of Chan-Paton charges, whereas $R_M$
parameterizes the gauge symmetry breaking, induced by the orbifold action.
For the M\"obius amplitude one has two options:
\[
{\cal M}^{(0)}_1\sim \frac{1}{2}I_M
({\hat V}_4 {\hat O}_4-{\hat S}_4 {\hat S}_4)\,,\quad
{\cal M}^{(0)}_2\sim -\frac{1}{2}I_M
({\hat O}_4 {\hat V}_4-{\hat C}_4 {\hat C}_4)
\]
where for the second M\"obius amplitude, we introduced a discrete
Wilson line. Inspection of the above M\"obius amplitudes 
requires the gauge group to be symplectic in the first
case and unitary in the second. Tadpole conditions give $I_M=16$ and $R_M=0$, 
which can be extracted from the transverse channel amplitudes
$\tilde{\cal K}^{(0)}$, $\tilde{\cal
M}^{(0)}$ and $\tilde{\cal A}^{(0)}$. 
This fixes the size of the gauge group.
In order to have a consistent particle interpretation of
${\cal A}^{(0)}+{\cal M}^{(0)}$ we have to choose
\begin{eqnarray*}
1.) & I_M=M_1+M_2\; ,\quad R_M=M_1-M_2\\
2.) & I_M=M+{\overline {M}}\; ,\quad R_M=i(M-{\overline M})
\end{eqnarray*}
The resulting spectrum in the first case comprises a vector in the
adjoint of ${\rm Sp}(8)^{\otimes 2}$ and a hypermultiplet
in the bifundamental representation $(8,8)$ and in the second
case a vector in the adjoint of ${\rm U}(8)$ and a hypermultiplets in 
$28\oplus{\overline {28}}$.

In the limit $R\to 0$, the amplitudes reveal that new tadpoles arise,
due to odd windings that become massless.
On the other hand, taking the limit $R\to \infty$ one recovers
the two six dimensional models with quantized
antisymmetric tensor background \cite{BS,KST,carlo} and gauge
group ${\rm Sp}(8)^{\otimes 4}$ with matter in the representations
$(8,8,1,1)$, $(8,1,8,1)$, $(1,8,1,8)$, and $(1,1,8,8)$ for the first case and
gauge group ${\rm U}(8)^{\otimes 2}$ with matter in the representations
$(28\oplus{\overline{28}},1)$, $(1,28\oplus{\overline{28}})$, 
$(8,\bar 8)$ and $(\bar 8,8)$ for the second case.

\begin{acknowledgments}
I am grateful to  C. Angelantonj and I. Antoniadis for collaboration
and interesting discussions. I would also like to thank the Organizers
of the Carg\`ese Summer School.
This work was supported in part by EEC TMR contract ERBFMRX-CT96-0090.
\end{acknowledgments}

\end{document}